# The bitwise operations in relation to the concept of set

Krasimir Yordzhev

Trakia University, Stara Zagora, Bulgaria

e-mail: krasimir.yordzhev@trakia-uni.bg

**ABSTRACT**

We contemplate this article to help the teachers of programming in his aspiration for giving some appropriate and interesting examples. The work will be especially useful for students-future programmers, and for their lecturers.

Some of the strong sides of these programming languages C/C++ and Java are the possibilities of low-level programming. Some of the means for this possibility are the introduced standard bitwise operations, with the help of which, it is possible directly operate with every bit of an arbitrary variable situated in the computer's memory.

In the current study, we are going to describe some methodical aspects for work with the bitwise operations and we will discuss the benefit of using bitwise operations in programming. The article shows some advantages of using bitwise operations, realizing various operations with sets.

*Keywords: Bitwise operation; Set; integer representation of sets, Class; Overloading of operators.*

## 1. INTRODUCTION

The use of bitwise operations is a powerful means during programming with the languages C/C++ and Java. In the current study, we are going to describe some *methodical aspects* for work with the bitwise operations (see also [1, 2, 3, 4]).

In the paper [4], we described an algorithm for receiving a Latin square of arbitrary order using operations with sets. Unfortunately, the programming languages C/C++ and Java do not support a standard type "set" [5], whereas for example the Pascal language does. For this reason, if there should be a need to use the operations with sets in the realization of some of our algorithms, we have to look for additional instruments to work with sets, such as, for example, the associative containers **set** and **multiset**, realized in *Standard Template Library (STL)* [6, 7, 8, 9]. We can also use the template class **set** of the system of computer algebra "Symbolic C++", which programming code is given in details in [10], or abstract class **IntSet**, that presents the interface of set realized through a dynamic array and ordered binary tree, described in [11]. Of course, another class set also can be built, and specific methods of this class can be described, as a means of training. This is a good exercise for students when the cardinality of the basic (universal) set is not very big. For example, the standard Sudoku puzzle has basic set the set of the integers from 1 to 9 plus the empty set.

The purpose of this paper is to show the advantages of bitwise operations to work with sets in the C++ programming language. This, of course, can be easily converted in the Java programming language, which has a similar syntax as in C++ [12, 13]. Here we will create own class **set** by describing specific methods for working with sets.

## 2. BITWISE OPERATIONS - BASIC DEFINITIONS, NOTATIONS AND EXAMPLES

Bitwise operations can be applied for integer data type only, i.e. they cannot be used for float and double types.

We assume, as usual that bits numbering in variables starts from right to left, and that the number of the very right one is 0.

Let x, y and z are integer variables or constants of one type, for which bits are needed. Let x and y are initialized (if they are variables) and let the assignment z = x & y; (*bitwise AND*), or z = x | y; (*bitwise inclusive OR*), or z = x ^ y; (bitwise exclusive OR), or z = ~x; (*bitwise NOT*) be made. For each $i = 0,1,2,…,w-1$, the new contents of the $i$-th bit in z will be as it is presented in the Table 1.

**Table 1. Bitwise operations in programming languages C/C++ and JAVA**

| $i$-th bit of x | $i$-th bit of y | $i$-th bit of z = x & y; | $i$-th bit of z = x \| y; | $i$-th bit of z = x ^ y; | $i$-th bit of z = ~x; |
|---|---|---|---|---|---|
| 0 | 0 | 0 | 0 | 0 | 1 |
| 0 | 1 | 0 | 1 | 1 | 1 |
| 1 | 0 | 0 | 1 | 1 | 0 |
| 1 | 1 | 1 | 1 | 0 | 0 |

In case that k is a nonnegative integer, then the statement z = x<<k (*bitwise shift left*) will fill the $(i+k)$-th bit of z the value of the $k$ bit of x, where $i = 0,1,…,w-k-1$, and the very right $k$ bits of x will be filled by zeroes. This operation is equivalent to a multiplication of x by $2^k$.

The statement z=x>>k (*bitwise shift right*) works the similar way. However, we must be careful if we use the programming language C or C++. In various programming environments this operation has different interpretations – somewhere $k$ bits of z from the very left place are compulsory filled by 0 (*logical displacement*), and elsewhere the very left $k$ bits of z are filled with the value from the very left (sign) bit (*arithmetic displacement*), i.e. if the number is negative, then the filling will be with 1. Therefore, it is recommended to use unsigned type of variables (if the opposite is not necessary) while working with bitwise operations. In the Java programming language, this problem is solved by introducing the two different operators: z=x>>k and z=x>>>k [12, 13].

Bitwise operations are left associative.

The priority of operations in descending order is as follows: ~ (*bitwise NOT*); the arithmetic operations * (*multiplication*), / (*division*), % (*remainder or modulus*); the arithmetic operations + (*addition*) - (*subtraction)*; the bitwise operations << and >>; the relational operations <, >, <=, >=, ==, !=; the bitwise operations &,^ and |; the logical operations && and ||.

Below we show some elementary examples of using the bitwise operations.

**Example 1:** To compute the value of the i-th bit (0 or 1) of an integer variable x we can use the function:

```
int BitValue(int x, unsigned int i) {
    int b = ((x & 1<<i) == 0) ? 0 : 1;
```

```
    return b;
}
```

**Example 2:** Directly from the definition of the operation bitwise shift left (<<) follows the efficiency of the following function computing $2^n$, where $n$ is a nonnegative integer:

```
unsigned int Power2(unsigned int n) {
    return 1<<n;
}
```

**Example 3:** The integer function $f(x) = x \% 2^n$ implemented using operation bitwise shift right (>>).

```
int Div2(int x, unsigned int n) {
    int s = x<0 ? -1 : 1;
   /* s = the sign of x */
    x = x*s;
 /* We reset the sign bit of x */
    return (x>>n)*s;
}
```

When we work with negative numbers we must consider that in the computer the presentation of the negative numbers is through the so called true complement code. The following function gives us how to code the integers in the memory of the computer we work with. For simplicity we are going to work with type short, but it is not a problem for the function to be overloaded for other integer types, too.

**Example 4:** A function showing the presentation of the numbers of type short in the memory of the computer.

```
void BinRepl(short n) {
    int b;
    int d = sizeof(short)*8 - 1;
    while (d>=0) {
       b= 1<<d & n ? 1 : 0;
       cout<<b;
       d--;
    }
}
```

In Table 2 we give some experiments with the function **BinRepl**:

**Table 2. Presentation of some numbers of type short in the memory of the computer**

| An integer of type short | Presentation in memory |
|---|---|
| 0 | 0000000000000000 |
| 1 | 0000000000000001 |
| -1 | 1111111111111111 |
| 2 | 0000000000000010 |
| -2 | 1111111111111110 |
| $16 = 2^4$ | 0000000000010000 |
| $-16 = -2^4$ | 1111111111110000 |
| $26 = 2^4+2^3+2$ | 0000000000011010 |
| $-26 = -(2^4+2^3+2)$ | 1111111111100110 |
| $41 = 2^5+2^3+1$ | 0000000000101001 |
| $-41 = -(2^5+2^3+1)$ | 1111111111010111 |
| $32767 = 2^{15} - 1$ | 0111111111111111 |
| $-32767 = -(2^{15} - 1)$ | 1000000000000001 |
| $32768 = 2^{15}$ | 1000000000000000 |
| $-32768 = -2^{15}$ | 1000000000000000 |

Compare the function presented in Example 4 to the next function presented in Example 5.

**Example 5:** A function that prints an integer in binary notation.

```
void DecToBin(int n) {
    if (n<0) cout<<'-';
/* Prints the sign - , if n<0: */
    n = abs(n);
    int b;
    int d = sizeof(int)*8 - 1;
    while ( d>0 && (n & 1<<d ) == 0 ) d--;
/* Skips the insignificant zeroes at the beginning: */
```

```
        while (d>=0) {
            b= 1<<d & n ? 1 : 0;
            cout<<b;
            d--;
        }
    }
```

**Example 6:1** The following function calculates the number of 1 in an integer $n$ written in a binary notation. Here again we ignore the sign of the number (if it is negative) and we work with its absolute value.

```
    int NumbOf_1(int n) {
    n = abs(n);
    int temp=0;
    int d = sizeof(int)*8 - 1;
        for (int i=0; i<d; i++)
            if (n & 1<<i) temp++;
        return temp;
    }
```

## 3. A PRESENTATION OF THE SUBSETS OF A SET

Let $M = \{\beta_0, \beta_1, \ldots, \beta_{m-1}\}$, $|M| = m$, be a finite set. Each subset of $M$ could be denoted by means of a Boolean vector $b(A) = \langle b_0, b_1, \ldots, b_{m-1}\rangle$, where $b_i = 1 \Leftrightarrow \beta_i \in A$ and $b_i = 0 \Leftrightarrow \beta_i \notin A$, $i = 0,1,2,\ldots,m-1$. As we proved in [14], a great memory economy could be achieved, if instead of boolean vectors, we use the presentation of the non-negative integers in a binary notation, where the integer 0 corresponds to empty set, while the integer $2^m - 1$, which in a binary notation is written by means of $m$ identities, corresponds to the basic set $M$. Thus, a natural one to one correspondence between the integers of the closed interval $[0, 2^m - 1]$ and the set of all subsets of $M$ is achieved. The integer $a \in [0, 2^m - 1]$ corresponds to the set $A \subseteq M$, if for every $i = 0,1,2,\ldots,m-1$ the $i$-th bit of the binary representation of $a$ equals 1 if and only if $\beta_i \in A$. In this way, the need of the use of bitwise operations naturally arises in cases involving the computer realization of various operations with sets.

Such an approach is comfortable and significantly effective when the basic set $M$ is with relatively small cardinal number $m = |M|$. A significant importance has also the operating system and programming environment that is used. This is so, because to encode a set, which is a subset of $M$, where $|M| = m$, with the above mentioned method $m$ bits are necessary. If $k$ bits are necessary for the integer type in the programming environment, then $\left\lfloor \frac{m}{k} \right\rfloor + 1$ variables of that certain type will be necessary, so as to put the above mentioned ideas into practice, where $\lfloor x \rfloor$ denotes the function "the whole part of $x$". For example, when $n \leq 5$, four bytes (thirty-two bits) are necessary to write a program that can solve a Sudoku puzzle in the size of $n^2 \times n^2$ if we use the set theory method [15]. In this case, every set of the kind $A = \{\alpha_1, \alpha_2, \ldots, \alpha_s\} \subseteq \{1,2,\ldots,n^2\}$ and the empty set could be simply encoded with an integer.

In particular, let $A \subseteq \{1,2,\ldots,n\}$. We denote by $\mu_i(A)$, $i = 1,2,\ldots n$ the functions

$$\mu_i(A) = \begin{cases} 1 & if \quad i \in A \\ 0 & if \quad i \notin A \end{cases} \tag{1}$$

Then we represent uniquely the set $A$ by the integer

$$v(A) = \sum_{i=1}^{n} \mu_i(A) 2^{i-1}, \quad 0 \leq v(A) \leq 2^n - 1, \tag{2}$$

where $\mu_i(A)$, $i = 1,2,...,n$ is given by formula (1). In other words, each subset of the set $\{1,2,...,n\}$, we will represent uniquely with the help of an integer from the interval $[0, 2^n - 1]$ (*integer representation of sets*).

It is readily seen that

$$v(\{1,2,...,n\}) = 2^n - 1. \tag{3}$$

Evidently if $A = \{a\}$, i.e. $|A| = 1$, then

$$v(\{a\}) = 2^{a-1}. \tag{4}$$

The empty set $\emptyset$ is represented by

$$v(\emptyset) = 0. \tag{5}$$

## 4. A PROGRAM IMPLEMENTATION OF SUBSETS OF THE SET {1, 2, ... , 32} USING THE BITWISE OPERATIONS

We consider the set
$$\mathcal{U} = \{1,2,...,32\},$$
which we call *basic*.

Here we will describe a class whose objects can be all subsets of $\mathcal{U}$, including the empty set. The class will contain a single field – an integer **n** of type **unsigned int**, the binary record of which will represent the considered set. Thus $k$-th bit of this record is 1 if and only if the integer $k+1$ belongs to the set represented by **n** (Bit numbering starts from zero). Methods of this class will be various operations with sets.

The class **Set_N**, which we create, will have two constructors. The first one has no parameters and initializes the empty set. The second one has one parameter – a nonnegative integer, the binary record of which determines the set. Thus, the empty set can be initialized in two ways – with no parameter or with a parameter equal to 0. In many programming environments, the basic set $\mathcal{U}$ is initialized with the standard constant Maxint, which in our case is equal to $2^{32} - 1$. Using the operation << (bitwise shift to the left), this constant can be calculated as shown in the following example:

**Example 7:**

    Set_N A, B(0);
    unsigned int mx = ((1<<31) - 1)*2 + 1;
    Set_N U(mx);

In Example 7, the sets A and B are initialized as empty sets in both different ways, and U is the basic set, i.e. U is the set containing all integers from 1 to 32.

Let the sets $A, B \subseteq \mathcal{U} = \{1,2,...n\}$, which will be the objects of the class we create and let the integer $k \in \mathcal{U}$. Consider the following operations with sets that will realize as methods of the class Set and which, by overloading some operators, will have their own suitable notations:

- The intersection $A \cap B$ of two sets. This operation we will denote with **A*B**.
- The union $A \cup B$ of two sets. This operation we will denote with **A+B**.
- The union $A \cup \{k\}$ of the set $A$ with the one-element set $\{k\}$. This operation we will denote with **A+k**.
- Adding the integer $k \in \mathcal{U}$ to the set $A$. This operation we will denote with **k+A**.

**Remark:** Here we have to note that from the algorithmic point of view A + k and k + A are realized differently, taking into account the standard of C++ programming language, regardless of commutativity for the operation of union of two sets.

- Removing the integer k from the set A. If $k \notin A$ then A does not change. This operation we will denote with **A-k**.
- Let $A \setminus B = \{k \mid k \in A \ \& \ k \notin B\}$. This operation we will denote with **A-B**.
- Checking whether $A \supseteq B$, that is, whether the set $A$ contains the proper subset $B$. This operation we will denote with **A>=B**. The result is true or false.
- Checking whether $A \subseteq B$, that is, whether the set $A$ is proper subset of the set $B$. This operation we will denote with **A<=B**. The result is true or false.
- Verifying that sets $A$ and $B$ are equal to each other we will denote with **A==B**. The result is true or false.
- Checking whether the sets $A$ and $B$ are different will be denoted by **A!=B**. The result is true or false.
- To verify that an integer $k \in \mathcal{U}$ belongs to the set $A \subseteq \mathcal{U}$, we will use the method (function) **A.in(k)**. The result is true or false.

Below we offer a specification the class **Set_N**:

```
class Set_N
{
/*
        The set is encoded by non-negative integer n in binary notation:
*/
        unsigned int n;
        public:
/*
        Constructor without parameter – creates empty set:
*/
        Set_N();
/*
        Constructor with parameter – creates a set containing the integer i, if and only if the i-th bit of the parameter k is 1:
*/
```

```cpp
        Set_N(unsigned int k);
/*
        Returns the integer n that encodes the set:
*/
        int get_n() const;
/*
        Overloading of the operators *, +, -, >=, <=, == and !=
*/
        Set_N operator * (Set_N const &);
        Set_N operator + (Set_N const &);
        Set_N operator + (unsigned int);
        friend Set_N operator + (unsigned int, Set_N const &);
        Set_N operator - (unsigned int);
        Set_N operator - (Set_N const &);
        bool operator >= (Set_N const &);
        bool operator <= (Set_N const &);
        bool operator == (Set_N const &);
        bool operator != (Set_N const &);
/*
        Checks whether the integer k belongs to the set:
*/
        bool in(unsigned int k);
/*
        Destructor
*/
        ~Set_N();
}
```

Below we describe a realization of the methods of class **Set_N**, with substantial use of bitwise operations:

```cpp
Set_N::Set_N()
{
        n = 0;
}
Set_N::Set_N(unsigned int k)
{
```

```cpp
        n = k;
}

int Set_N::get_n()
{
        return n;
}

Set_N Set_N::operator * (Set_N const &s)
{
        return (this->n) & s.get_n();
}

Set_N Set_N::operator + (Set_N const &s)
{
        return (this->n) | s.get_n();
}

Set_N Set_N::operator + (unsigned int k)
{
        return (this->n) | (1<<(k-1));
}

Set_N operator + (unsigned int k, Set_N const &s)
{
        return (1<<(k-1)) | s.get_n();
}

Set_N Set_N::operator - (unsigned int k)
{
        int temp = (this->n) ^ (1<<(k-1));
        return (this->n) & temp;
}

Set_N Set_N::operator - (Set_N const &s)
```

```cpp
{
    int temp = this->n ^ s.get_n();
    return (this->n) & temp;
}

bool Set_N::operator >= (Set_N const &s)
{
    return (this->n | s.get_n()) == this->n;
}

bool Set_N::operator <= (Set_N const &s)
{
    return (this->n | s.get_n()) == s.get_n();
}

bool Set_N::operator == (Set_N const &s)
{
    return ((this->n ^ s.get_n()) == 0);
}

bool Set_N::operator != (Set_N const &s) {
    return !((this->n ^ s.get_n()) == 0);
}

bool Set_N::in(int k)
{
    return this->n & (1<<(k-1));
}
```